\documentstyle[12pt]{article}
\newcommand{\bi}{{\bar\i}}
\newcommand{\bj}{{\bar\j}}

\begin{document}

\bigskip
\begin{flushright}
NSF-PHY97-22022\\
hep-th/0202118
\end{flushright}
\bigskip
\bigskip

\centerline{\Large \bf D3-brane Action in a Supergravity Background:} 
\medskip
\centerline{\Large \bf the Fermionic Story }
\bigskip
\bigskip
\bigskip
\centerline{\bf Mariana Gra\~na}
\bigskip
\centerline{Department of Physics}
\centerline{University of California}
\centerline{Santa Barbara, CA 93106}
\centerline{\it mariana@physics.ucsb.edu}
\bigskip
\bigskip
\bigskip
\bigskip
\begin{abstract}
\bigskip

Using the $\kappa$-symmetric action for a D3-brane, we study the interaction between its world-volume fermions and a bosonic type IIB supergravity background preserving 4-dimensional Lorentz invariance. We find that the renormalizable terms in the action include only coupling between the fermions and the 3-form flux in the combination $(*_6 G_{(3)}-iG_{(3)})$, which is zero for a class of supersymmetric and nonsupersymmetric solutions. We also find the magnetic and electric dipole moments for the fermions, which are proportional to the derivative of the dilaton-axion. We show that different gauges to fix the $\kappa$-symmetry give the same interaction terms, and prove that these terms are also $SL(2,{\bf R})$ self-dual. We interpret our results in terms of ${\mathcal N}=1$ supersymmetric gauge theory on the D-brane.   

\end{abstract}

\newpage

\section{Introduction}

Since the realization made by Polchinski in 1995 that D-branes carry Ramond-Ramond charges \cite{pol}, they have been considerably studied (see for example \cite{dbraneprimer} for a review). Neverthe- less, we still lack of a full understanding of the action governing their dynamics.   

The low energy effective action of a bosonic Dp-brane is the sum of the Born-Infeld and Wess-Zumino terms. These describe the dynamics of the bosonic massless fields on the D-brane in a general supergravity background, when the fields vary slowly compared to the string scale. The supersymmetrization of this action has been worked out by different groups \cite{cederwall3-brane} - \cite{APS}. The supersymmetric action looks exactly like the bosonic one, but in the former case the brane lives in superspace, and every space-time field is promoted to a superfield. The explicit form of the action is given in terms of the background supervielbein and super NS-NS and R-R gauge fields. Such as it is, it's difficult to extract information about the dynamics of the world-volume fermionic fields with the purpose of studying D-brane phenomenology. 

Explicit fermionic dependence to all orders in the fermions have been obtained in flat space in \cite{cederwall3-brane} and \cite{APS1}, and in $AdS_5 \times S^5$ in  \cite{MT}.  In this paper we wish to obtain the explicit coupling of a type IIB supergravity background with nontrivial dilaton-axion and both 5 and 3-form fluxes to the world-volume fermions on a D3-brane, to second order in the latter. This is important for any understanding of brane-world models both in supersymmetric and nonsupersymmetric backgrounds. It is also indispensable if we wish to know the long range fields created by a particular D3-brane polarization which, as we will see, has 3-form flux besides the known self-dual 5-form flux.  

An expansion of space-time superfields in terms of component fields for a general background was carried out in \cite{IIAgaugecompl}, following a method known as gauge completion, proposed in \cite{11dgaugecompl}. This made it possible to obtain the explicit linear coupling between world-volume fermions and a IIA supergravity background for a D0-brane. The action obtained this way was checked by an independent method relating it to the matrix description of M-theory. We will perform the corresponding expansion for the IIB case.   

In fact, we will obtain the action for the world-volume fermions by three different methods: from the known renormalizable 4-dimensional actions with ${\mathcal N}=1$ supersymmetry supplemented by 10-dimensional covariance and the relationship found in \cite{GP1} between the 3-form flux and the field theory superpotential; from a direct calculation along the lines of \cite{11dgaugecompl}, and from T-duality of the D0-brane action worked out in \cite{IIAgaugecompl}. The actions obtained by the three different methods agree, providing a nice and useful result for future use. 
   
The rest of the paper is organized as follows: in Section 2 we find the possible couplings between world-volume fermions and background fields from 10-dimensional covariance, 4-dimensional Lorentz invariance and gauge invariance. In Section 3 we find these couplings for a supersymmetric background corresponding to N D3-branes plus a 3-form flux perturbation, worked out in \cite{GP1}. We also show that any smooth deformation of this background lies in a particular class of solutions studied in \cite{GP1}-\cite{BVFLM}. In Section 4 we perform the expansion of the superfields in terms of component fields, and find thereof the desired interaction terms from the supersymmetric D3-brane action. The results obtained are checked in Section 5 with the action obtained by T-dualizing the D0-brane action. In Section 6, we interpret the results in terms of the 4-dimensional field theory on the D3-brane. We state our conclusions in Section 7, and show in the Appendix the conventions chosen and the supergravity equations used.        

\section{Action from 10-dimensional covariance}

We want to find the renormalizable terms in the action that represent the coupling between the fermionic fields on a D3-brane and a bosonic 
type IIB background. The massless bosonic fields of the type IIB superstring theory consist of the dilaton $\phi$, the metric tensor $G_{MN}$ and the antisymmetric 2-tensor
$B_{MN}$ in
the NS-NS sector, and the axion $C$, the 2-form potential $C_{MN}$, and the
four-form field $C_{MNPQ}$ with self-dual five-form field strength in the
R-R sector.  Their fermionic superpartners are a complex Weyl
gravitino $\psi_{M}$ ($\Gamma^{10}\psi_{M}=-\psi_{M}$) and a complex Weyl
dilatino $\lambda$ ($\Gamma^{10}\lambda=\lambda$). The theory has $D=10$,
$\mathcal
{N}$=2 supersymmetry with two Majorana-Weyl supersymmetry parameters of the same chirality
$\varepsilon_{1,2}$ $(\Gamma^{10}\varepsilon_{1,2} =-\varepsilon_{1,2})$. These two spinors are usually combined in the supergravity literature into a complex Weyl spinor.
The two scalars $C$ and $\phi$ can be combined into a complex field $\tau = C+
ie^{-\phi}$ which parameterizes the $SL(2,{\bf R})/ U\left(1\right)$ coset space.

We are interested in backgrounds that preserve Lorentz invariance on the world-volume of the D3-brane, which lies in the 0123 coordinates (henceforth called $\mu,\nu...$). Consequently, we will assume a background geometry of the warped form:

\begin{equation}
ds^{2} = Z^{-1/2}\eta_{\mu\nu}dx^{\mu}dx^{\nu}+ Z^{1/2}
\widetilde{ds}{}_6^2\ ,
\label{metricZ} 
\end{equation}
where the 6-dimensional space is a complex manifold ${\it M}$, labeled in each patch by complex coordinates $z^i$, $i=1,2,3$.

To preserve $SO(3,1)$ invariance, the 5-form field strength has to be:
\begin{equation}
F_5=d\chi_4+ *\, d\chi_4 \,\ , \qquad\ \chi_4=f(x^m)\,\, dx^0\wedge dx^1\wedge dx^2\wedge dx^3
\label{F5}
\end{equation}
where $f$ is a function of the orthogonal coordinates $x^m$. There will be generically nonzero NS-NS and R-R 3-form fluxes orthogonal to the D3-brane.

The low energy fermionic degrees of freedom on the world-volume are the massless open string Ramond states, forming 
a 10-dimensional Majorana-Weyl spinor $\Theta$. From 10-dimensional covariance and gauge invariance, we can guess the form of the interaction between $\Theta$ and the background fields. From gauge invariance, we know that the field strengths and not the potentials should be involved in the Lagrangian. The R-R 3-form field strength $F_3$ and the NS-NS $H_3$ will be combined in the complex 3-form $G_{(3)}$:
\begin{equation}
G_{(3)} = F_{(3)} - \tau
H_{(3)}
\label{G3}
\end{equation}
which can have all $(3,0)$, $(2,1)$, $(1,2)$ and $(0,3)$ pieces with respect to the complex structure defined on ${\it M}$. This field might couple to $\Theta$ through a term like
\begin{equation}
\overline{\Theta} \Gamma^{mno} \Theta\, G_{mno}
\label{Gcoupling}
\end{equation}
We expect the axion and dilaton to appear in the form:
\begin{equation}
\overline{\Theta} \Gamma^{m} \Theta\, \partial_m \tau
\label{taucoupling}
\end{equation}
and the 5-form field strength as:
\begin{equation}
\overline{\Theta} \Gamma^{MNOPQ} \Theta\, F_{MNOPQ}
\label{F5coupling}
\end{equation}
where we have used capital letter indices to indicate that these run over the whole 10-d space. We can already rule out two of these interactions, namely (\ref{taucoupling}) and (\ref{F5coupling}), from the anticommutation relations and the Majorana property of $\Theta$. For a 32-component Majorana-Weyl spinor, the only nonzero fermion bilinears are 
\begin{equation}
\overline{\Theta} \Gamma^{MNP} \Theta \qquad      \overline{\Theta} \Gamma^{MNPQRST} \Theta
\label{only3}
\end{equation}
so we are left only with the possibility outlined in (\ref{Gcoupling}) of the world-volume fermions directly coupling to $G_{(3)}$.    

The Majorana-Weyl spinor $\Theta$ in the ${\bf \overline{16}}$ of $SO(9,1)$ splits 
into $({\bf\bar{2}},{\bf4})+({\bf 2},{\bf\bar{4}})$ of $SO(3,1)\otimes SO(6)$. This means that there are four 4-dimensional spinors, three fermions $\lambda^i$ in the ${\mathcal N}=1$ chiral multiplet and a gaugino $\psi$ in the ${\mathcal N}=1$ vector multiplet.  The 32 x 32 Dirac gamma matrices can be 
decomposed as
\begin{equation}
\Gamma^{\mu}=\gamma^\mu \otimes 1 \, , \qquad\    \Gamma^{m}= \gamma_{(5)} \otimes \gamma^m
\end{equation}
where $\gamma^{\mu}$ and $\gamma^m$ are Dirac gamma matrices corresponding to $SO(3 ,1)$ and $SO(6)$ respective- ly. For the $SO(6)$ matrices, it is convenient to group them into complex holomorphic and antiholomorphic ones. The spinors are eigenstates of $S_i = \gamma^{i}\gamma^{\bi}-\frac{1}{2}$ with eigenvalues $\pm \frac{1}{2}$ . The fermions $\lambda ^i$ are built with $SO(6)$ spinors that have positive (negative) $S_i$ eigenvalue and negative (positive) $S_j, j\neq i$ for the ${\bf 4}$ (${\bf \overline{4}}$) representation. The gaugino is built from the spinor that has all positive (negative) $S_i$ eigenvalues. 

There are two independent $SO(3,1)\otimes SO(6)$ invariants that we can build out of $G_{(3)}$, namely the one in Eq.(\ref{Gcoupling}) and another one with a power of $\gamma_{(5)}$ inserted. Then, in 4-dimensional terms, the coupling to the background $G_{(3)}$ field is:
\begin{eqnarray}
a\,\overline{\Theta} \gamma_{(5)}\Gamma^{mno}\Theta \, G_{mno}+ b\,\overline{\Theta} \Gamma^{mno}\Theta \, G_{mno}\, \sim \,&& (a+b)\left(\psi \psi G_{ijk} + \lambda^i \lambda^j S_{ij} + \lambda^i \psi G_{ij}\;^j\right)\nonumber\\
 &+& (a-b) \left(\bar{\psi} \bar{\psi} G_{\bi\bj \bar{k}}+ \bar{\lambda}^{\bi} \bar{\lambda}^{\bj} S_{\bi\bj} + \bar{\lambda}^{\bi} \bar{\psi} G_{\bi\bj}\;^{\bj}\right) 
\label{4dG3coupling}
\end{eqnarray}
where $S_{ij}$ involves the $(1,2)$ piece of $G_{(3)}$ in the symmetric combination:
\begin{equation}
S_{ij}=\frac{1}{2}\left(\hat{\epsilon}_{ikl}G_{j\bar{k}\bar{l}}+ \hat{\epsilon}_{jkl}G_{i \bar{k} \bar{l}}\right)
\end{equation}
and $\hat{\epsilon}$ is the orthogonal frame epsilon, with values $\pm1$. 

For backgrounds that do not preserve Lorentz invariance there would be additional pieces in the couplings (\ref{4dG3coupling}), namely those where the 3-form flux has one, two or three indices along the D3-brane. We have restricted ourselves to Lorentz-invariance preserving backgrounds, so we will not see the effect of those possible terms.   

The gaugino mass is then proportional to the $(3,0)$ piece of $G_{(3)}$, the mass matrix for the fermions in the chiral multiplet is given by the $(1,2)$ piece, and the coupling between them by the nonprimitive $(2,1)$. In the next section we will interpret this in terms of $\mathcal{N}$ = 1 supersymmetry.

\section{Interaction Lagrangian from $\mathcal{N}$ = 1 supersymmetry}

 A perturbation expansion in powers of $G_{(3)}$ around the black D3-brane solution $(\ref{metricZ})$-$(\ref{F5})$ was carried out in $\cite{GP1}$, with $\widetilde{ds}{}_6^2$ the flat metric, $f=-\frac{1}{4Z}$ and a constant dilaton. In order to preserve $\mathcal N$=1 supersymmetry, the first order 3-form flux, has to obey
\begin{equation}
G_{ijk}=G_{ij}\,^j=0 \, , \qquad\ S_{ij}=Z\,\partial_i\partial_j W(z^i)
\label{G3susy}
\end{equation}
where $W$ is a holomorphic function proved to be proportional to the superpotential that breaks $\mathcal N$= 4 into $\mathcal N$= 1 in the dual field theory. The complex coordinates $z^i$ on which $W$ depends, are in the gauge theory the scalar components $\phi^i$ of the three chiral multiplets.  

This result can be generalized by carrying out a first order perturbation in all the bosonic fields, keeping the metric as in (\ref{metricZ}) with $\widetilde{ds}{}_6^{2}$ being flat plus a first order correction. 

The supersymmetry variations of the dilation and gravitino in the Einstein frame are  
\cite{Schwarz} \footnote{In this Section we use Einstein frame quantities, and a complex spinor $\varepsilon$, according to the conventions that are mostly used in supergravity literature.}. 
\begin{equation}
\delta\lambda=- \frac{e^{\phi}}{2}\Gamma^{M}\partial_{M}\tau \, \varepsilon^* +
\frac{e^{\frac{1}{2}\phi}}{24}\Gamma^{MNP}G_{MNP}\, \varepsilon
\label{dilatino} 
\end{equation}

\begin{equation}
\delta\psi_{M}=(D_{M}+i\frac{e^{\phi}}{4}\,\partial_M C)\, \varepsilon + \frac{i}{480} 
\Gamma^{M_{1}...M_{5}}
F_{M_{1}...M_{5}}\Gamma_M\varepsilon+i\frac{e^{\frac{1}{2}\phi}}{96}\left(\Gamma_{M}^{PQR}G_{PQR}-
9\Gamma^{PQ}G_{MPQ} \right) \varepsilon^{*}
\label{gravitino}
\end{equation}
where $D_{M}$ is the covariant derivative with respect to the metric $g_{MN}$.
 
The first order dilatino variation enforces $\tau$ to be holomorphic and sets $G_{(3)}$ to obey the first of the two conditions in ($\ref{G3susy}$). The first order gravitino variation splits into two equations. One of them is exactly the same as in \cite{GP1}, so the first order 3-form flux perturbation obeys all the conditions found there. The important ones for us are those in (\ref{G3susy}). 
The other one is
\begin{equation}
\left(D_{M}+i\frac{e^{\phi}}{4}\partial_M C\right)^{(1)} \varepsilon^{(0)} + \frac{i}{480} 
\left(\Gamma^{M_{1}...M_{5}}
F_{M_{1}...M_{5}}\Gamma_M\right) ^{(1)} \varepsilon^{(0)} = 0
\label{fograv}
\end{equation}
where the subscripts (0) and (1) denote the order in the perturbation expansion. The first order product of $F_{(5)}$ and gamma matrices involve the fist order perturbation in $F_{(5)}$ as well as that of the metric. The supersymmetry of a configuration of D3/D7-branes, which has nontrivial holomorphic $\tau$ that depends only on one of the complex coordinates, was worked out in \cite{GP2}. In that case, the transverse metric has to be of the form
\begin{equation}
\widetilde{ds}{}_6^{2}=2\Bigl(
dz^{1}d\bar{z}^{1}+ dz^{2}d\bar{z}^{2}+ e^{-\phi(z)} dz \,d\bar{z} \Bigr)\
\label{metric}
\end{equation}

If we consider a smooth deformation of this background, we can subtract off this perturba- tion from (\ref{fograv}) and set $\tau$ to a constant. We can also subtract off a perturbation of the 5-form field. Any such perturbation that preserves Lorentz invariance has to be of the form (\ref{F5}), and the result of adding it it's just a perturbation of the warp factor proportional to $f$. Thus, we can also consider the 5-form flux and the warp factor as being unperturbed. Then, insisting that the supersymmetry parameter shouldn't depend on the longitudinal coordinates, the $\mu$ component of (\ref{fograv}) tells us that
\begin{equation}
F_{M_1 ... M_5}^{(0)} \left(\Gamma^{M_1 ... M_5} \right)^{(1)}=0
\end{equation}
This is the same that appears in the $m$ component of (\ref{fograv}). This means that from the first term in this equation, we finally find that the first order covariant derivative of $\epsilon^{(0)}$ has to be zero, which implies that the perturbed metric has to be Ricci flat. In summary, any smooth deformation of the D3/D7 system has to lie in the class studied in \cite{GP1} (Eq.(\ref{G3susy}) for $G_{(3)}$ and a metric perturbation of the "warped Calabi-Yau" form).

Going back to the interaction between the world-volume fermions and the background, Eq.(\ref{4dG3coupling}), we see that in the $\mathcal N$= 1 "perturbed D3/D7-brane background", the gaugino remains massless and there is no gaugino-$\lambda^i$ fermion interaction. On the other hand, these fermions in the chiral multiplet acquire a mass proportional to the second derivative of the holomorphic function $W$. This fits nicely with $\mathcal N$= 1 supersymmetry in 4-dimensions, where the mass of the chiral fermions is just the second derivative of the superpotential. We will follow this discussion in Section 6, after we have found the exact interaction Lagrangian.

\section{Couplings from $\kappa$-symmetric D3-brane action}

The supersymmetric extension of the Dirac-Born-Infeld and Wess-Zumino actions for a Dirichlet p-brane was worked out in \cite{cederwall3-brane}-\cite{Bergshoeff} in terms of supersymmetric space-time variables. It looks like the ordinary bosonic action. For a D3-brane in the string frame, it is
\begin{equation}
S=-\mu_3\int d^4\zeta e^{- \Phi}\sqrt{-det \left({\bf g}_{ij} + F_{ij}-{\bf B}_{ij}\right)}+\mu_3\int e^{F -{\bf B}} \wedge {\bf C}
\label{susyaction}
\end{equation}
The use of boldface indicates that the space-time field has been promoted to a superfield. Here 
\begin{equation}
{\bf C}=\oplus_n {\bf C}_{(n)}
\end{equation}
is the collection of all RR potentials pulled back to the world-volume.

The brane is embedded in superspace, labeled by the coordinate $Z^M=(x^m,\theta^{\mu})$ \footnote{In this and the following Sections, we use capital letter subindices to indicate bosonic as well as fermionic indices. $a,b,...$ refer to bosonic tangent space indices; $m,n,...$ are bosonic coordinate space indices; $i,j,...$ are world-volume bosonic indices and Greek letters refer to fermionic indices.}. The ${\mathcal N}=2$ 10-dimensional supersymmetry is realized as $\delta \theta = \epsilon,\, \delta x^m=\bar{\epsilon} \Gamma^m \theta$. The space-time fermions $\theta$ are a pair of real sixteen-component Majorana-Weyl spinors. The IIB fermionic index $\mu$ has to be understood as the product of a Majorana-Weyl index and an $SO(2)$ index (see Appendix A for more detail). 

To project a space-time index -bosonic or fermionic- into the world-volume we must contract with the coordinate differential $dZ^M$. Taking for example ${\bf B}_{(2)}$, the projection amounts to
\begin{equation}
B_{ij}= \partial_i Z^M \partial_j Z^N {\bf B}_{MN} \nonumber 
= \partial_i x^m    
\partial_j x^n {\bf B}_{mn} + 2 \partial_{[i} x^m \partial_{j]} \theta ^{\mu} 
{\bf B}_{m\mu} + \partial_i \theta^{\mu}\partial_j \theta ^{\nu}{\bf B}_{\mu\nu}
\label{finalsplit}
\end{equation}

The superspace constraints for the superfields are solved in \cite{HoweWest}. Here, we will use a different approach. We will find the expansion of every space-time field in powers of $\theta$ up to second order, using a method known as gauge completion. This method was developed for 11-dimensional supergravity in \cite{11dgaugecompl} and applied to IIA in the Einstein frame in \cite{IIAgaugecompl}. We will apply the same techniques in order to obtain the type IIB case in the string frame. For an alternative way to obtain the expansion of the vielbein, see \cite{normalcoordinates}. 

The idea of gauge completion is to compare the superspace diffeomorphisms of a given superfield with the supersymmetry transformation of the corresponding component field. A superspace diffeomorphism acts like
\begin{equation}
\delta {\bf C}_{MN}= \Sigma^P\partial_P {\bf C}_{MN}+2\partial_{[M} \Sigma^P {\bf C}_{P|N]}
\label{superdiffeo}
\end{equation}
where $\Sigma$ is the superdiffeomorphism parameter. To zeroth order in $\theta$, it is
\begin{equation}
\Sigma^m = \zeta^m\, , \qquad\ \Sigma^{\alpha}=\varepsilon^{\alpha}
\end{equation}
where $\zeta^m$ and $\epsilon^{\alpha}$ are the usual diffeomorphism and supersymmetry transformation parameters. The space-time superfields are at zeroth order
\begin{eqnarray}
{\bf e}_m^a&=&e_m^a \nonumber \\
{\bf e}_m^{\alpha}&=&\psi_m^{\alpha} \nonumber \\
{\bf e}_{\mu}^a&=&0  \nonumber \\
{\bf e}_{\mu}^{\alpha}&=&\delta_{\mu}^{\alpha} \\
{\Phi}&=&\phi \nonumber \\
{\bf B}_{mn}&=&B_{mn} \nonumber \\
{\bf B}_{m\mu}&=&0 \, \nonumber \\
&\vdots& \nonumber
\label{zerothorder}
\end{eqnarray}
where "..." indicates that all superfields with bosonic indices have as zeroth order component the corresponding bosonic field, and all the superfields with at least one fermionic index are zero. 

To obtain the first order terms, let us illustrate with the dilaton superfield  $\Phi$, to which we apply a superdiffeomorphism. This amounts to an ordinary diffeomorphism plus a supersymmetry transformation. The latter has to be equal to the supersymmetry transforma- tion of the bosonic field $\phi$, i.e.
\begin{equation}
\Sigma ^{\alpha} \partial_{\alpha} \Phi= \varepsilon^{\alpha}\partial_{\alpha}{\Phi} \equiv \delta_{\epsilon}\phi={\overline \varepsilon} \lambda
\end{equation}
which implies
\begin{equation}
\Phi=\phi+\theta \lambda + {\mathcal O}(\theta ^2)
\end{equation}

The same can be done for the vielbein and the gauge potentials. For these, the second term in (\ref{superdiffeo}) doesn't contribute, except when performing a superdiffeomorphism to ${\bf e}_m^{\alpha}$. In that case we have
\begin{eqnarray}
\delta {\bf e}_m^\alpha &=& \Sigma ^{\beta} \partial_{\beta }{\bf e}_m^{\alpha } 
+\partial_ m \varepsilon ^{\beta } {\bf e}_{\beta }^{\alpha } \nonumber \\
&=& \varepsilon ^{\beta }\partial_{\beta } {\bf e}_m^{\alpha }+\partial_m \varepsilon ^{\alpha } \nonumber \\
&\equiv& \delta e_m^{\alpha }=\delta \psi_m^{\alpha }=\partial_m \varepsilon ^{\alpha }+ ...
\end{eqnarray}  
so the derivative of the supersymmetry parameter cancels, and the first order vielbein with mixed components is equal to the supersymmetry variation of the gravitino without this derivative. 

To obtain the fermionic components of the fields we need to know $\Sigma $ to first order in $\theta$. To that end, we perform the commutator of two superdiffeomorphisms on any superfield, say $\Phi$, and make it equal to the commutator of two supersymmetries, i.e.
\begin{equation}
\delta_{[\Sigma_1, \Sigma_2]} \Phi =\left(\varepsilon_1^{\alpha }\partial_{\alpha }\Sigma^a_2\right)\partial_a \Phi
+ \left( \varepsilon _1^{\alpha } \partial _{\alpha }\Sigma _2^{\beta }\right) \partial _{\beta } \Phi + ... - 1 \leftrightarrow 2
\label{commutator}
\end{equation}
where ... indicates terms that involve the ordinary diffeomorphism parameter $\zeta$. This should be equal to the commutator of two supersymmetries applied to the component field $\phi $, which gives a combination of a diffeomorphism and a gauge transformation in the case of a gauge field, or a Lorentz transformation in the case of the vielbein. We are interested in the coordinate transformation, whose parameter is $\zeta^a= {\overline \varepsilon}_2 \Gamma ^a \varepsilon _1$. The second term in (\ref{commutator}) is zero for a bosonic background, while the first one is the coordinate transformation. We have then
\begin{equation}
\Sigma ^a=\zeta^a- \frac{1}{2}{\overline \theta} \Gamma ^a \varepsilon + {\mathcal O}(\theta ^2)
\end{equation}

Now we can compute the first order term in the expansion of ${\bf e}_{\mu}^a$ by performing a superdiffeomorphism 
\begin{eqnarray}
\delta {\bf e}_{\mu }^a &=& \Sigma ^{\beta } \partial _{\beta }{\bf e}_{\mu }^a + \partial _{\mu }\Sigma ^b {\bf e}_b^a+...\nonumber \\
&=& \varepsilon ^{\beta }\partial _{\beta } {\bf e}_{\mu }^a- \frac{1}{2}({\overline \varepsilon} \Gamma^a)_{\mu }
\end{eqnarray}
this implies
\begin{equation}
{\bf e}_{\mu }^a= \frac{1}{2} (\overline{\theta} \Gamma ^a)_{\mu}
\end{equation}

In the same way we can obtain ${\bf B}_{m\mu }$, ${\bf C}_{m\mu }$ and ${\bf C}_{mno\mu }$, including in these cases the gauge transformations that arise when performing two supersymmetry transformations on a given gauge field (see appendix A for the form of the gauge parameters).  

To get the action to second order, we just need the second order expansion of the superfields with all bosonic components, and the first order expansion of the superfields with one fermionic index. The second order terms are obtained by the same argument as the first order ones. 

We will use static gauge, in which the coordinates that label the world-volume $\zeta^i$ are identified with the space-time coordinates $x^i$. In that gauge, the space-time spinor $\theta$ becomes a spinor in the world-volume. The superspace generalization of the action possesses a local fermionic symmetry called $\kappa$-symmetry \cite{ksymm}. This symmetry allows us to remove half of the fermions by a gauge choice, making the number of fermionic degrees of freedom in the world-volume equal to the bosonic ones. We will work in a general gauge, where the two space-time Majorana-Weyl $\theta_1$ and $\theta_2$ are related to the world volume spinor $\Theta$ by

\begin{equation}
\theta_1=a\Theta \qquad\ \theta_2=b\Theta, \qquad\ a^2+b^2=1
\label{ab}
\end{equation}

Then the only nonzero fermion bilinears are those containing antisymmetric products of three or seven gamma matrices (cf Eq(\ref{only3})) with either $\sigma^1$, $\sigma^3$ or no Pauli matrices multiplying, so many of the terms that appear to second order in the expansion of the superfields vanish. Inside a fermion bilinear, $\sigma^1$ will be equal to $2ab$ and $\sigma^3$ to $a^2-b^2$.  

Up to the order that we need in $\Theta $ (second in the case of a bosonic component, and first in the case of fermionic), we get, for a bosonic background
\begin{eqnarray}
{\bf e}_m^a&=&e_m^a + \frac{1}{8} \overline{\Theta} \Gamma^{abc} \Theta w_{mbc}-  \frac{1}{16} (a^2-b^2)  \overline{\Theta} \Gamma^{anp} \Theta H_{mnp}+ \frac{1}{32} 2ab\, e^{\phi} \overline{\Theta} \Gamma^{np}\,_m \Theta F'^{a}\,_{np} \nonumber\\
& & + \frac{1}{32} 2ab\, e^{\phi} \overline{\Theta} \Gamma^{anp} \Theta F'_{mnp}-\frac{1}{96} 2ab\, e^{\phi} e^a_m \overline{\Theta} \Gamma^{npq} \Theta F'_{npq}       \nonumber \\
{\bf e}_{\mu }^a&=& \frac{1}{2} (\overline{\Theta}\Gamma^a)_{\mu} \nonumber \\   
{\bf e}_m^{\mu}&=& 0 \nonumber \\
{\bf e}_{\mu}^{\alpha}&=& \delta_{\mu}^{\alpha} \nonumber \\
\Phi&=& \phi- \frac{1}{48} (a^2-b^2)  \overline{\Theta} \Gamma^{pqr} \Theta H_{pqr}-\frac{1}{48} 2ab\, e^{\phi}\overline{\Theta} \Gamma^{mnp} \Theta F'_{mnp}+    \nonumber \\
{\bf B}_{mn} &=& B_{mn} +\frac{1}{4}(a^2-b^2)\overline{\Theta} \Gamma^{ab}\,_{[m} \Theta \,w_{n]ab} -\frac{1}{8} 2ab\, e^{\phi} \overline{\Theta} \Gamma_{mn}\,^p \Theta \partial_{p}C+ ... \nonumber\\
{\bf B}_{m\mu}&=& \frac{1}{2}(a^2-b^2) (\overline{\Theta} \Gamma_m)_{\mu} \nonumber\\
{\bf C}_{mn}&=& C_{mn}-\frac{1}{4} 2ab\, e^{-\phi}  \overline{\Theta} \Gamma^{ab}\,_{[m} \Theta \,w_{n]ab}+\frac{1}{8}(a^2-b^2)\overline{\Theta} \Gamma_{mn}\,^{p} \Theta \,\partial_{p}C +\frac{1}{4} 2ab\, \overline{\Theta} \Gamma_{mn}\,^{p} \Theta \partial_{p}\phi + C B_{mn}|_{\Theta^2}+... \nonumber\\
{\bf C}_{m\mu}&=& C B_{m\mu} -\frac{1}{2}2ab\, e^{-\phi}\left(\overline{\Theta} \Gamma_m\right)_{\mu} \nonumber\\
{\bf C}_{mnop}&=& C_{mnop}+ \frac{1}{48} (a^2-b^2) \overline{\Theta} \Gamma_{mnop}\,^{qrs} \Theta F'_{qrs}+\frac{1}{48} 2ab\, e^{-\phi} \overline{\Theta} \Gamma_{mnop}\,^{qrs} \Theta H_{qrs}+  ... \nonumber \\
{\bf C}_{mno\mu}&=&0+...
\label{expansionIIB} 
\end{eqnarray}
where the $+...$ in the second order ${\bf B}_{mn}$, ${\bf C}_{mn}$  and ${\bf C}_{mnop}$ and first order ${\bf C}_{mno\mu}$ are terms that are zero for our background, when projecting this fields to the world-volume. 

To second order in $\Theta $, the action (\ref{susyaction}) for a background with 3-form flux orthogonal to the D3-brane, is
\begin{eqnarray}
\frac{S}{\mu_3}=-\int d^4\zeta e^{-\phi} (1-\Phi|_{\Theta^2})\left\{-det [g_{ij}+ 2\partial_{(i}\Theta ^{\mu} \partial_{j)}\,x^b {\bf e}_{\mu}^a|_{\Theta } \eta_{ab}+ 2 \partial_i x^m \partial_j x^b {\bf e}_{m}^a|_{\Theta^2} \eta_{ab} \right. \nonumber \\
+  \left. F_{ij}- {\bf B}_{ij}|_{\Theta^2} +2 \partial_{[i} \Theta^{\mu} {\bf B}_{j]\mu}]\right\}^{\frac{1}{2}} \nonumber \\
+ \frac{1}{4!} \hat{\epsilon}^{ijkl} \int d^4\zeta \left[ C_{ijkl}+{\bf C}_{ijkl}|_{\Theta^2} 
+ 6 F_{[ij} \left( {\bf C}_{kl]} |_{\Theta^2} - 2\partial_{[k} \Theta^{\mu} {\bf C}_{l]\mu} 
 - C ({\bf B}_{kl} |_{\Theta^2} -2\partial_{[k}\Theta^{\mu}{\bf B}_{l]\mu})  \right)\right]
\label{actionexp}
\end{eqnarray}
where $\hat{\epsilon}$ is the Levi-Civita tensor in the orthogonal frame, with values $\pm 1$.

Inserting the expansions (\ref{expansionIIB}) in (\ref{actionexp}) and using, for the DBI part:
\begin{equation}
\sqrt{det \left( 1+M \right)}=1+\frac{1}{2}tr M -\frac{1}{4} trM^2 +\frac{1}{8} (tr M)^2 +{\cal O}(M^3)
\end{equation}
we get, to second order in $\Theta $ and up to terms proportional to $F$, changing the background metric in (\ref{metricZ}) to the string frame, 
\begin{eqnarray}
\frac{1}{\mu_3}{\mathcal L}_{DBI}= {\mathcal O}(\Theta ^0)+Z^{-1}\left(- \frac{1}{2}\overline{\Theta} \Gamma^iD_i\Theta - \frac{1}{48} (a^2-b^2)\overline{\Theta }\Gamma^{pqr}\Theta H_{pqr} +\frac{1}{48} 2ab\, e^{\phi}\overline{\Theta }\Gamma^{pqr}\Theta F'_{pqr} \right. \nonumber\\
\left. -\frac{1}{2} (a^2-b^2) F^{ij}\, \overline{\Theta} \Gamma_iD_j\Theta +\frac{1}{16} 2ab\,  e^{\phi} F^{ij}\, \overline{\Theta} \Gamma_{ij}\,^{m}\Theta \partial_m C \right) +{\mathcal O}(\Theta^4, F^2)
\label{Ldbid3}
\end{eqnarray}
where ${\mathcal O}(\Theta ^0)$ are the usual bosonic terms.

For the Wess-Zumino Lagrangian, we get
\begin{eqnarray}
\frac{1}{\mu_3}{\mathcal L}_{WZ}= {\mathcal O}(\Theta ^0)+ Z^{-1} \left( \frac{1}{48} (a^2-b^2) e^{\phi} \overline{\Theta }\Gamma^{pqr}\Theta (*_6 F'_{(3)})_{pqr} + \frac{1}{48} 2ab\, \overline{\Theta }\Gamma^{pqr}\Theta (*_6 H_{(3)})_{pqr} \right. \nonumber\\
\left. -\frac{1}{2} 2ab\, (*F)^{ij}\, \overline{\Theta} \Gamma_iD_j\Theta + \frac{1}{16}(a^2-b^2) e^{\phi} (*F)^{ij}\, \overline{\Theta }\Gamma_{ij}\,^{m}\Theta \partial_m C +\frac{1}{8} 2ab\, (*F)^{ij}\, \overline{\Theta }\Gamma_{ij}\,^{m}\Theta \partial_m \phi \right)
\label{Lwzd3}
\end{eqnarray}

Collecting (\ref{Ldbid3}) and (\ref{Lwzd3}), we get a D3-brane Lagrangian for the world-volume fermions

\begin{eqnarray}
\frac{1}{\mu_3}{\mathcal L}&=& {\mathcal O}(\Theta ^0)+ Z^{-1}\left(-\frac{1}{2}\overline{\Theta} \Gamma^i D_i \Theta + \frac{1}{48} e^{\phi} \overline{\Theta }\Gamma^{pqr}\Theta Re\left[(a+ib)^2\left(*_6 G- i G\right)_{pqr}\right]\right. \nonumber \\
&-& \frac{1}{2}(a^2-b^2) F^{ij}\, \overline{\Theta} \Gamma_iD_j\Theta -\frac{1}{2} 2ab\, (*F)^{ij}\, \overline{\Theta} \Gamma_iD_j\Theta +\frac{1}{16}2ab\, e^{\phi} F^{ij}\, \overline{\Theta} \Gamma_{ij}\,^m\Theta \partial_m C \nonumber\\
&+& \left. \frac{1}{16}(a^2-b^2) e^{\phi} (*F)^{ij}\, \overline{\Theta }\Gamma_{ij}\,^{m}\Theta\, \partial_m C + \frac{1}{8} 2ab\, (*F)^{ij}\, \overline{\Theta }\Gamma_{ij}\,^{m}\Theta \,\partial_m \phi \right)
\label{intercoupling}
\end{eqnarray}

The world-volume fermions don't couple to the 5-form flux (or to the derivative of the warp factor) as we suggested in Section 2, and they couple to the derivative of the dilation-axion only through the field strength. As far as the coupling to the 3-form flux, they only couple to the combination $(*_6 G_{(3)}-iG_{(3)})$. In the class of exact supersymmetric solutions to IIB supergravity studied in \cite{GP1, Gubser}, which includes the Klebanov-Strassler solution for the deformed conifold \cite{KS}, this combination is zero. In a more general class of supersymmetric and nonsupersymmetric solutions studied in \cite{GKP}, where the localized sources obey a sort of BPS condition, the 3-form flux $G_{(3)}$ must also be imaginary self-dual.

The couplings found in (\ref{intercoupling}) indicate that certain polarizations of the D3-brane have magnetic H-dipole moment, as well as magnetic D5-brane dipole moment. This in turn implies that some polarizations of the D3-brane will produce a generically nonzero value of the 3-form flux. In \cite{IIAgaugecompl}, the long range supergravity fields were worked out for the different polarizations of a D0-brane. Having found the couplings between the background and the world-volume fermions to second order in the latter, we can obtain the long range supergravity fields produced by a D3-brane up to order $r^{-5}$.

\subsection*{Gauge invariance and $SL(2,{\bf R})$ duality}

The action found in (\ref{intercoupling}) is both gauge invariant and $SL(2,{\bf R})$ self-dual (when written in the Einstein frame). Let's first show this for the term involving the coupling to the three-form flux. In the Einstein frame (see Eqs.(\ref{transf1}) and (\ref{transf2}) below), this coupling is
\begin{equation}
\frac{1}{48}e^{\frac{1}{2}\phi} Re\left[(a+ib)^2\left(*_6 G- i G\right)_{pqr}\right] \,
\overline{\Theta }\Gamma^{pqr}\Theta 
\label{gcouplingeinstein}
\end{equation}

Under an $SL(2,{\bf R})$ transformation
\begin{equation}
\tau'=\frac{a\tau+b}{c\tau+d}, \quad \quad \quad \pmatrix{H'_{(3)} \cr F\,'_{(3)}\cr}= \pmatrix{d & c \cr b & a\cr} \pmatrix{H_{(3)} \cr F_{(3)}\cr}\
\end{equation}
the combination appearing in (\ref{gcouplingeinstein}) picks up a phase
\begin{equation}
e^{\frac{1}{2}\phi'}G\,'_{(3)}=e^{-i\alpha} e^{\frac{1}{2}\phi}G_{(3)}, \qquad e^{i\alpha}=\frac{c\tau+d}{|c\tau+d|}
\label{S-duality}
\end{equation}
Also under gauge transformations this combination picks up a phase
\begin{equation}
(a'+ib')^2=e^{i \beta} (a+ib)^2
\label{gaugered}
\end{equation}

Any one of these phases can be removed by a chirality rotation in the fermion $\Theta$
\begin{equation}
\Theta '= e^{i \delta \gamma_{(5)}} \Theta
\end{equation}
whose action on (\ref{gcouplingeinstein}) is
\begin{equation}
Re\left[\left(*_6 G- i G\right)_{pqr}\right] \,
\overline{\Theta}'\,\Gamma^{pqr}\Theta '= 
Re\left[e^{-2i \delta} \left(*_6 G- i G\right)_{pqr}\right] \,
\overline{\Theta}\,\Gamma^{pqr}\Theta 
\end{equation}
Setting $\delta=\frac{\beta}{2}$ or $\delta=-\frac{\alpha}{2}$ we can absorb the phases that appear when we change the gauge or perform an $SL(2, {\bf R})$ transformation. This part of the action is then both gauge and $SL(2,{\bf R})$ invariant.

The terms that involve the field strength in (\ref{intercoupling}) are also gauge invariant and self-dual. In the Einstein frame, using again the equations below (\ref{transf1}) and (\ref{transf2}), these are
\begin{eqnarray}
\frac{1}{{\mu_3}}{\mathcal L} _{(Einstein)}=&&{\mathcal O}(\Theta ^0)+ {\mathcal O}(F^0,\Theta ^2)\nonumber\\
&+& Z^{-1}\left(-\frac{1}{2}(a^2-b^2)e^{-\frac{1}{2}\phi} F^{ij}\, \overline{\Theta} \Gamma_iD_j\Theta -\frac{1}{2} 2ab\, e^{-\frac{1}{2}\phi} (*F)^{ij}\, \overline{\Theta} \Gamma_iD_j\Theta \right. \nonumber\\
&+& \left. \frac{1}{16}2ab\, e^{\frac{1}{2}\phi} F^{ij}\, \overline{\Theta} \Gamma_{ij}\,^m\Theta \partial_m C + \frac{1}{16}(a^2-b^2) e^{\frac{1}{2}\phi} (*F)^{ij}\, \overline{\Theta }\Gamma_{ij}\,^{m}\Theta\, \partial_m C \right. \nonumber\\
&-& \left. \frac{1}{16} (a^2-b^2) e^{-\frac{1}{2} \phi} F^{ij}\, \overline{\Theta }{\Gamma_{ij}}^{m}\Theta \,\partial_m \phi + 
\frac{1}{16} 2ab\, e^{-\frac{1}{2}\phi} (*F)^{ij}\, \overline{\Theta }\Gamma_{ij}\,^{m}\Theta \,\partial_m \phi \right)
\label{Fcouplingeinstein}
\end{eqnarray}
where the terms proportional to the derivative of the dilaton come in part from translating to the Einstein frame the term in (\ref{intercoupling}) containing the covariant derivative of the fermion coupled to the field strength.     

The first term linear in the filed strength is zero from 
\begin{eqnarray}
\overline{\Theta } \Gamma^{i} D^{j} {\Theta} \partial_{j}A_{i}&=& \frac{1}{2}\overline{\Theta } \Gamma^{i} \{ \Gamma^{j}, \not\! D \}\Theta \partial_{j}A_{i} \nonumber\\
&=& \left(\overline{\Theta } D^{i} \Gamma^{j} {\Theta}-\frac{1}{2}\overline{\Theta } \not\! D \Gamma^{i} \Gamma^{j} {\Theta}\right)  \partial_{j} A_{i}
\end{eqnarray}
where in the second equality we used the equation of motion in a fermion bilinear $\overline{\Theta} \Gamma^{i} D_{i}\Theta=0$. Moving the first term to the left hand side, we get the combination that appears in the action, which is zero after integrating the second term by parts and using the Bianchi identity $dF=0$. The same can be shown for the second term, involving $*F$.

The remaining four terms in (\ref{Fcouplingeinstein}) can be combined in the form
\begin{equation}
\frac{1}{16} 
Im \left[ (a+ib)^2 (F+i*F)^{ij} \frac{\partial_m \tau}{\sqrt{\tau_2}}\right]\, \overline{\Theta }\Gamma_{ij}\,^{m}\Theta 
\label{Fcouplingeinsteincompact}
\end{equation}

This is obviously invariant under $C'=C+1$, and we will show that it is also invariant under an S-duality $\tau'=-\frac{1}{\tau}$, along the lines of Seiberg and Witten's prove of S-duality in ${\mathcal N}=2$ super Yang Mills \cite{SW}. Introduce a Lagrange multiplier vector field $A_D$, add to the Lagrangian a term $\frac{1}{2}(*F_D)^{ij}F_{ij}$ and integrate out $F$. The result of doing this is the usual bosonic term with $F$ replaced by $F_D$ and $\tau$ replaced by $-\frac{1}{\tau}$, and for the term involving the fermions, we get  
\begin{equation}
\frac{1}{16} Im \left[ (a+ib)^2 (F_D+i*F_D)^{ij} \frac{\partial_m \tau}{\tau\sqrt{\tau_2}}\right]\,\overline{\Theta }\Gamma_{ij}\,^{m}\Theta 
\end{equation}
But this is just what we would get if we change $\tau$ in (\ref{Fcouplingeinsteincompact}) by $-\frac{1}{\tau}$, up, again,  to a phase transformation, i.e.
\begin{equation}
\frac{\partial_m \tau}{\sqrt{\tau_2}}\,\,\,\overrightarrow{\scriptstyle{\tau \rightarrow -\frac{1}{\tau}}}\,\,\, \frac{\partial_m \tau}{\tau\sqrt{\tau_2}}e^{-i\alpha}
\end{equation}
where $\alpha$ was defined in (\ref{S-duality}), and in the present case $c=1, d=0$. A gauge redefinition also amounts to adding a phase, as in (\ref{gaugered}). Any two of these phases can now be absorbed by a duality rotation of the world-volume field strength.  

We have thus shown that the action is gauge invariant, and is invariant under $\tau\rightarrow \tau+1$ and $\tau \rightarrow -\frac{1}{\tau}$, making it $SL(2,{\bf R})$ self-dual. It is nice to see how both symmetries act in a similar way, adding a phase that can be removed by a redefinition of a world-volume field. 

\section{T-duality of D0-brane Action}

The interactions between world-volume fermions and the supergravity background for a D0-brane to order $\Theta^2$ have been worked out in the Einstein frame in \cite{IIAgaugecompl}. The explicit form of the action is shown in the weak field limit, keeping only terms linear in the background field. But it is not hard to get the action including nonlinear terms, as well as terms involving time derivatives of $\Theta$, that they set to zero. It is 	
\begin{eqnarray}
S = &-& \mu_0 \int d\tau e^{-\frac{3}{4}\phi} (1- \frac{3}{4}\Phi |_{\Theta^2}) \left((- g_{MN}- 2e_{M a}{\bf e}^a_{N}|_{\Theta^2}) \dot{x}^{M} \dot{x}^{N}
- 2 \dot{\Theta }^{\mu} {\bf e}_{\mu}^a \eta_{ab} \dot{x}^b \right)^{\frac{1}{2}} \nonumber \\
&+& \mu_0\int d\tau \left(C_M + {\bf C}_M|_{\Theta^2}\right) \dot{x}^M +   \dot{\Theta}^{\mu}C_{\mu}|_{\Theta }   
\label{action}
\end{eqnarray}
where boldface fields are superfields, and the indices $M,N$ are bosonic indices that run over time and space. 

The expansions found in \cite{IIAgaugecompl} for a bosonic background, in the gauge $\Gamma^{11}\Theta=-\Theta$ are
 \begin{eqnarray}
{\bf e}_m^a&=& e_m^a +\frac{i}{8}\overline{\Theta} \Gamma^{abc}\Theta w_{mbc} +\frac{i}{64} e^{-\frac{1}{2}\phi}\overline{\Theta}\Gamma_m\,^{op}\Theta H^a\,_{op}+\frac{3i}{64} e^{-\frac{1}{2}\phi} \overline{\Theta}\Gamma^{ano}\Theta H_{mno} \nonumber \\
&&-\frac{i}{192} e^{-\frac{1}{2}\phi}e^a_m\overline{\Theta}\Gamma^{nop}\Theta H_{nop} \nonumber \\
{\bf e}_{\mu}^a&=&\frac{i}{2}(\overline{\Theta} \Gamma^a)_{\mu} \nonumber \\
{\bf e}_m^{\mu}&=& \frac{1}{64}e^{\frac{3}{4}\phi} (\Gamma_m\,^{no}\Theta)^{\mu} F_{no}-\frac{5}{192} e^{\frac{1}{4}\phi} (\Gamma^{nop} \Theta)^{\mu} F'_{mnop} \nonumber \\
{\bf e}_{\mu}^{\nu}&=& \delta_{\mu}^{\nu} \nonumber \\
\Phi&=& \phi + \frac{i}{48}e^{-\frac{1}{2}\phi} \overline{\Theta}\Gamma^{mnp}\Theta H_{mnp} \nonumber \\
{\bf B}_{mn}&=&  -\frac{i}{4}e^{\frac{1}{2}\phi} \overline{\Theta}\Gamma_{[m}\,^{op}\Theta\, w_{n]op} -\frac{3i}{32} \overline{\Theta}\Gamma_{[m}\,^{op}\Theta H_{n]op}- \frac{i}{8}e^{\frac{1}{2}\phi} \overline{\Theta}\Gamma_{mn}\,^{p}\Theta \,\partial_{p}\phi    \nonumber \\
{\bf C}_{m}&=& C_m -\frac{i}{16}\overline{\Theta}\Gamma_m\,^{no}\Theta F_{no} -\frac{i}{48} e^{-\frac{1}{2}\phi} \overline{\Theta}\Gamma^{nop}\Theta F'_{mnop} \nonumber \\
{\bf C}_{\mu}&=&-\frac{i}{2}e^{-\frac{3}{4}\phi}(\overline{\Theta} \Gamma^{11})_{\mu}
\label{expansionsIIA}
\end{eqnarray}

Inserting (\ref{expansionsIIA}) in (\ref{action}) and expanding the square root in the DBI part we get, to first order in the velocity and second order in $\Theta $ \footnote{We have assumed that the metric splits into a time-time component and space-space components. This must hold if we want the T-dual metric to look like (\ref{metricZ}).}  
\begin{eqnarray} 
&&\frac{1}{\mu_0 \sqrt{-g_{00}}} {\mathcal L}_{DBI} = {\mathcal O}(\Theta ^0)\nonumber \\
&+&\frac{i}{48} e^{-\frac{5}{4}\phi} \overline{\Theta} \Gamma^{mnp} \Theta H_{mnp} -\frac{i}{16} e^{-\frac{5}{4}\phi} \overline{\Theta} \Gamma^{0mn} \Theta H_{0mn} -  \frac{i}{8} e^{-\frac{3}{4}\phi} \overline{\Theta} \Gamma^{0mn} \Theta w_{0mn} 
- \frac{i}{2} e^{-\frac{3}{4}\phi} \overline{\Theta } \Gamma^{0} \dot{\Theta} \nonumber\\
&+& \left( -\frac{i}{4} e^{-\frac{3}{4}\phi} \overline{\Theta } \Gamma_{(m|}\,^{np} \Theta w^{|0)}\,_{np} -  \frac{i}{16} e^{-\frac{5}{4}\phi} \overline{\Theta } \Gamma_{(m|}\,^{np} \Theta H^{|0)}\,_{np}     
-\frac{i}{2} e^{-\frac{3}{4}\phi} g^{00} \overline{\Theta } \Gamma_m \dot{\Theta} \right) \dot{x}^m 
\label{Ldbief}
\end{eqnarray}
where the indices $m,n,...$ run over spatial directions only.

The Wess-Zumino part of the action gives 
\begin{eqnarray} 
\frac{1}{\mu_0}{\mathcal L}_{WZ}= {\mathcal O}(\Theta ^0)
&-&\frac{i}{16} \overline{\Theta} \Gamma_0\,^{mn} \Theta F_{mn} -\frac{i}{48}
e^{-\frac{1}{2}\phi} \overline{\Theta } \Gamma^{mnp} \Theta F'_{0mnp}\nonumber\\ 
&-&\left( \frac{i}{16} \overline{\Theta } \Gamma_m\,^{np} \Theta F_{np} +\frac{i}{48} e^{-\frac{1}{2}\phi} \overline{\Theta } \Gamma^{npq} \Theta F'_{mnpq}\right) \dot{x}^m
\label{Lwzsf}
\end{eqnarray} 

This result is given in the Einstein frame, and we would like to change it to the string frame. We expect to have an overall factor of $\sqrt{g_{00 (string)}}e^{-\phi}$ in front of the Lagrangian. Using $g_{mn (Einstein)}= e^{-\frac{1}{2}\phi }g_{mn (string)}$, which implies 
\begin{eqnarray}
\sqrt{-g_{(4)(Einstein)}}= e^{-\phi }\sqrt{-g_{(4)(string)}} ,\qquad \Gamma^m_{(Einstein)}= e^{\frac{1}{4}\phi }\Gamma^m_{(string)} \nonumber \\ w^0\,_{mp(Einstein)}=\frac{1}{2}\,\partial_{[\hat{m}}\phi\, e^0_{\hat{p}](string)}+w^0\,_{mp(string)}
\label{transf1}
\end{eqnarray} 
-where a hat is used to remind that the indices are tangent space indices- we get that the fermion $\theta$ should transform as 
\begin{equation}
\theta_{(Einstein)}= e^{-\frac{1}{8}\phi }\theta_{(string)}
\label{transf2}
\end{equation}
This agrees with the rescaling of the spinors found in \cite{APPS} and also with the way the supersymmetry parameter $\varepsilon$ transforms \cite{HO}. Every RR field also transforms as $F_{(n)(Einstein)}=e^{-\phi} F_{(n)(string)}$.

In the string frame, the D0-brane Lagrangian is then 
\begin{eqnarray} 
&&\frac{1}{\mu_0 \sqrt{-g_{00}}} {\mathcal L}_{DBI (string)}= {\mathcal O}(\Theta ^0)\nonumber \\
&+& e^{-\phi}\left( \frac{i}{48} \overline{\Theta} \Gamma^{mnp} \Theta H_{mnp} -\frac{i}{16} \overline{\Theta} \Gamma^{0mn} \Theta H_{0mn} - \frac{i}{8} \overline{\Theta} \Gamma^{0mn} \Theta w_{0mn} 
- \frac{i}{2} \overline{\Theta } \Gamma^{0} \dot{\Theta}\right) \nonumber\\
&+& e^{-\phi}\left( -\frac{i}{4}\overline{\Theta } \Gamma_{(m|}\,^{np} \Theta w^{|0)}\,_{np} -  \frac{i}{8} \overline{\Theta } \Gamma_{(m|}\,^{np} \Theta H^{|0)}\,_{np}     
-\frac{i}{2} g^{00} \overline{\Theta } \Gamma_m \dot{\Theta} \right) \dot{x}^m 
\label{Ldbi}
\end{eqnarray}
for the DBI part, and
\begin{eqnarray} 
\frac{1}{\mu_0}{\mathcal L}_{WZ (string)}= {\mathcal O}(\Theta ^0)
&+&e^{-\phi}\left( \frac{i}{16} \overline{\Theta} \Gamma_0\,^{mn} \Theta F_{mn} -\frac{i}{48}
 \overline{\Theta } \Gamma^{mnp} \Theta F'_{0mnp}\right) \nonumber\\
&+& e^{-\phi}\left( -\frac{i}{16} \overline{\Theta } \Gamma_m\,^{np} \Theta F_{np} -\frac{i}{48} \overline{\Theta } \Gamma^{npq} \Theta F'_{mnpq}\right) \dot{x}^m
\label{Lwz}
\end{eqnarray} 
for the Wess-Zumino.

We want to T-dualize this action in the 1,2 and 3 directions to get the D3-brane action. If we implement T-duality in n directions labeled by $i,j,...$ the NS-NS and R-R fields transform as (see for example \cite{dbraneprimer})
\begin{eqnarray}
\tilde{{\cal G}}_{ab}&=& {\cal G}_{ab} -{\cal G}_{ai}{\cal G}^{ij}{\cal G}_{jb} \nonumber \\
\tilde{{\cal G}}_{ai}&=& {\cal G}^{ij}{\cal G}_{aj} \nonumber \\
\tilde{{\cal G}}_{ia}&=& -{\cal G}_{ja}{\cal G}^{ji} \nonumber \\
\tilde{{\cal G}}_{ij}&=& {\cal G}^{ij} \nonumber \\
\tilde{\phi}&=& \phi - \frac{1}{2}ln\, det({\cal G}_{ij}) \nonumber\\
\tilde{{\cal C}}^{q}_{a_1...a_{q-k}i_1...i_k}&=&\frac{1}{(n-k)!} \epsilon^{i_1...i_n} {\cal C}^{(n+q-2k)}_{a_1...a_{q-k}i_{k+1}...i_n} \nonumber \\
\tilde{A}^i&=& x^i
\label{T-dualityrules}
\end{eqnarray}
where the fields with tilde are the T-dual fields, $A^i$ is the world-volume gauge field, and we have defined  
\begin{eqnarray} 
{\cal G}_{ab} &=& g_{ab} - B_{ab} \nonumber \\
{\cal C}^{(n)} &=& C^{(n)} - C^{(n-2)} \wedge B +\frac{1}{2!} C^{(n-4)} \wedge B \wedge B +...
\end{eqnarray}

We want our T-dual metric to be of the form (\ref{metricZ}) multiplied by $e^{\frac{1}{2}\phi}$ to change it to the string frame, and the NS-NS and R-R T-dual 2-form fields to be orthogonal to the 3-brane. From (\ref{T-dualityrules}) we see that $B_{(2)}$ has to be originally orthogonal to it, and $g^{ij}=e^{\frac{1}{2}\phi}Z^{-\frac{1}{2}}\eta_{ij}$. 

Let's start with the velocity-independent terms in the DBI action (\ref{Ldbi}). The first one T-dualizes to 

\begin{equation}
\frac{i}{48} \sqrt{-g_{00 (string)}} e^{-\phi} \overline{\Theta } \Gamma^{mnp} \Theta H_{mnp} \rightarrow 
\frac{i}{48} Z^{-1} \overline{\Theta} (\Gamma^{mnp} H_{mnp} - 6 \Gamma^{iab} w_{iab}) \Theta 
\label{Hdual}
\end{equation}
     
This, together with the third and fourth terms in (\ref{Ldbi}), give the covariant derivative of $\Theta $ along the directions of the D3-brane: $\frac{i}{2}\overline{\Theta} \Gamma^i D_i \Theta$; $i=0,1,2,3$. The second term in (\ref{Ldbi}) vanishes when T-dualizing, by the condition of Lorentz invariance. 

As far as the T-dual of the velocity-independent terms in the Wess-Zumino Lagrangian (\ref{Lwz}), the first one T-dualizes to zero according to the Lorentz-invariance condition, and in the second one:
\begin{equation}
F'_{0mnp}=C_{0[mn,p]}-C_0 H_{mnp} \rightarrow -\frac{1}{3!} \epsilon^{ijk} C_{0ijk[mn,p]}-\frac{1}{3!} \epsilon^{ijk} C_{0ijk} H_{mnp}
\end{equation}
where in the first equality we imposed the condition that the fields are time independent, and $H_{(3)}$ doesn't have a time component.
Using, in form notation, 
\begin{equation}
dC_{(6)}= -e^{\phi} * (F_{(3)}-C\,H_{(3)}) + C_{(4)}\wedge H_{(3)}
\end{equation}
and combining it with the first term in (\ref{Hdual}), coming from the DBI Lagrangian, we get  
\begin{eqnarray}
-\frac{i}{48}
e^{-\phi} \overline{\Theta } \Gamma^{mnp} \Theta F'_{0mnp} +\frac{i}{48} \sqrt{-g_{00 (string)}} e^{-\phi} \overline{\Theta } \Gamma^{mnp} \Theta H_{mnp} \rightarrow \nonumber\\
- \frac{i}{48} e^{\phi} Z^{-1} \overline{\Theta } \Gamma^{mnp} \Theta Re \left[(*_6 G)_{mnp} - i G_{mnp}\right] 
\label{TdualGcoupl}
\end{eqnarray}

The terms proportional to the velocity, when T-dualizing, will give terms proportional to the world-volume field strength. So the terms we have just found are all the renormalizable terms on the world-volume. Collecting them, we get from T-duality that the renormalizable terms in the D3-brane Lagrangian \begin{equation}
{\mathcal L}_{D3}= {\cal O}(\Theta^0) + Z^{-1}\left(-\frac{i}{2}\overline{\Theta} \Gamma^i D_i \Theta - \frac{i}{48} e^{\phi} \overline{\Theta } \Gamma^{mnp} \Theta Re \left[(*_6 G)_{mnp} - i G_{mnp}\right]\right) 
\end{equation}
where $i$ runs fron 0 to 3.

This interaction between the world-volume fermions and the 3-form flux is exactly as found in the previous section (\ref{intercoupling})\footnote{The difference in factors of $i$ come from the $\Gamma$ matrices, which are taken to be real in Section V, while in \cite{IIAgaugecompl} they were chosen to be imaginary.} in the gauge $(a,b)=(1,0)$, providing a check of our results.

As for the terms involving the D0-brane velocity in the DBI Lagrangian (\ref{Ldbi}), the first and the last one combine to give a covariant derivative of the spinor, that T-dualizes to
\begin{equation}
-\frac{i}{2} e^{-\phi} \sqrt{g^{00}_{(string)}}\, \overline{\Theta } \Gamma_i D_0\Theta\, \dot{x}^i\rightarrow \frac{i}{2}Z^{-1} \overline{\Theta } \Gamma_i D_0 {\Theta} F^{i0}
\end{equation}
as appears in the D3-brane calculation (\ref{intercoupling}). The second one gives zero for our background.

In the Wess-Zumino action (\ref{Lwz}), the first term proportional to the velocity is zero after performing T-duality, and the second goes to
\begin{equation}
-\frac{i}{48} e^{-\phi} \overline{\Theta } \Gamma^{mnp} \Theta F'_{imnp} \dot{x}^i \rightarrow \frac{i}{16} Z^{-1} \overline{\Theta } \Gamma^{ijm} \Theta \partial_m C(*F)_{ij}
\end{equation}
This term is also in complete agreement with those found in the previous Section, also in the gauge $(a,b)=(1,0)$, concluding our check of the world-volume fermionic action for a D3-brane. 

From our results, it seems that the gauge choice $\Gamma^{11} \Theta=-\Theta$ in the IIA case corresponds to the choice $(a,b)=(1,0)$ (Eq. (\ref{ab})) in the IIB case. Anyway, we showed in the D3-brane case that the action is gauge invariant. 

\section{Gauge Theory on the D3-brane}

Splitting the spinor $\Theta$ in 10-dimensions into a 4-dimensional gaugino and three fermions $\lambda^i$ in the chiral multiplet as we did in Section 2, we can interpret our results in terms of gauge theory on the D3-brane. 

From the kinetic term for $\Theta $ we get the kinetic terms for the gaugino and $\lambda^i$ fermions\footnote{In this Section, since we are dealing with ordinary space, we go back to the conventions used at the beginning, where Greek subindices label directions along the D3-brane.}
\begin{equation}
-\frac{1}{2}Z^{-1}\left(\overline{\lambda}^{\bi} \overline{\sigma}^{\mu}\partial_{\mu} \lambda^i+ \overline{\psi} \overline{\sigma}^{\mu} \partial_{\mu} \psi\right)
\end{equation}
where the covariant derivative reduced to an ordinary derivative, since
\begin{equation}
\overline{\Theta} \gamma^{\mu} w_{\mu}\,^{ab}\Gamma_{ab}\Theta =-\frac{1}{2}\partial_m ln Z\, \overline{\Theta} \Gamma^m \Theta=0  
\end{equation}
from (\ref{only3}). As far as the interaction terms, in the first place, from the combination $(*_6 G_{(3)}-iG_{(3)})$ that we get in the renormalizable part of the action, only $G_{ijk}, S_{ij}$ and $G_{ij}\,^j$ survive. So, from (\ref{intercoupling}) and (\ref{4dG3coupling}), the renormalizable fermion-fermion interactions in the D3-brane are
\begin{equation}
\frac{Z^{-1}}{48}e^{\phi}\left(\lambda^i \lambda^j Re(S_{ij}) + \lambda^i \psi Re(G_{ij}\;^j)+ \psi \psi Re(G_{ijk}) \right)
\label{result4d}
\end{equation}
These components of the 3-form flux are in general functions of the coordinates $z^i$, which are in the gauge theory the scalar partners of $\lambda^i$. When the D-brane is located at some $z^i$, these scalar fields acquire a vev, giving masses to the fermions. 
 
In low energy effective ${\mathcal N}=1$ supersymmetric theories, these terms are constructed from the superpotential and the auxiliary fields $F$ and $D$ in the chiral and vector multiplet respectively 
\begin{equation}
{\mathcal L}_{{\mathcal N}=1}= -\frac{1}{2}\lambda^i \lambda^j \frac{\partial^2W}{\partial \phi^i \partial \phi^j}+\frac{i}{\sqrt{2}}\,D\,\frac{\partial\tau}{\partial\phi^i}\lambda^i\psi- F^i\,\frac{\partial\tau}{\partial \phi^i} \psi \psi +...
\label{Lsusy}
\end{equation}
where $+...$ stands for kinetic terms, and terms involving the gauge field.    

The term involving the mass of the fermions in the chiral multiplet fits nicely, as we mentioned in Section 3, with the result obtained in \cite{GP1}, where the symmetric combination of $(1,2)$ components of $G_{(3)}$ was found to be proportional to the superpotential in the field theory. 

The tree level gaugino mass is associated to F-term supersymmetry breaking. Comparing Eqs.(\ref{result4d}) and (\ref{Lsusy}), we see that this breaking comes in the 10-dimensional theory from the $(3,0)$ piece of the 3-form flux. Going back to (\ref{dilatino}), we see indeed that if the dilaton is holomorphic, a nonzero $(3,0)$ piece of $G_{(3)}$ breaks all the supersymmetry.

Finally, the coupling between the matter fermions and the gaugino, is associated with D-term supersymmetry breaking. Again, comparing with (\ref{result4d}), we see that this breaking comes from the nonprimitive $(2,1)$ 3-form flux, whose nonzero vacuum expectation value also breaks supersymmetry in the 10-dimensional theory.

Supersymmetry breaking in the field theory is mediated in this scenario by bulk supergravity fields. For other mechanisms of supersymmetry breaking in brane-world models see \cite{BWSB} and references therein.

The other terms in (\ref{intercoupling}), which have one power of the field strength, give in 4-dimensions
\begin{equation}
\overline{\Theta }\Gamma^{\mu\nu m}\Theta\, (F+i*F)_{\mu\nu}\,\partial_m \tau \sim \epsilon^{\bi\bj \bar{k}}  \lambda^i \overline{\sigma}^{\mu} \sigma^{\nu}\lambda^j (F+i*F)_{\mu\nu}\,\partial_{\bar{k}}\tau +  \lambda^i \overline{\sigma}^{\mu}\sigma^{\nu} \psi (F+i*F)_{\mu\nu}\,\partial_{i}\tau  
\end{equation}

These are transition electric and magnetic moments for the fermions in the chiral multiplet. In general, with multiple D3-branes, those fermions would be the quarks and leptons of the gauge theory. Considering the case of the neutrinos, transition moments have been speculated to be a possible explanation to the solar neutrino problem (see \cite{solarneutrino} and references therein). Models including neutrino transition magnetic moments both in the Standard Model \cite{neutrinosm} -including a fourth generation neutrino- and in the softly broken MSSM \cite{neutrinoMSSM} -keeping an unbroken $l_e - l_{\mu}$ symmetry- have been built. 

\section{Conclusions}

We have found the coupling between the world-volume fermions on a D3-brane and an $SO(3,1)$-invariant IIB supergravity background, to second order in the fermions. These only couple directly to the NS-NS and R-R 3-form flux, which implies that a single D3-brane in certain states has $H_{(3)}$ and $F_{(3)}$-dipole moments. The 3-form flux appears in the combination $(*_6 G_{(3)}-i G_{(3)})$, which is closely related to the equations of motion of type IIB supergravity. There are particular classes of supersymmetric and nonsupersymmetric exact solutions where this combination vanishes, rendering the background transparent to the brane (the bosonic world-volume degrees of freedom don't couple either to the 3-form flux when it is orthogonal to the brane). We have also found that the world-volume gauge field couples linearly only to the derivative of the dilaton-axion. We proved gauge invariance and $SL(2,{\bf R})$ duality of the action, noting that gauge redefinitions and $SL(2,{\bf R})$  transformations act in a fashion that can be offset by redefinitions of world-volume fields. 

By splitting the $SO(9,1)$ spinor $\Theta$ into representations of $SO(3,1) \otimes SO(6)$, we interpreted our results in terms of 4-dimensional gauge theories. The spinor $\Theta$ splits into a Weyl spinor in the ${\mathcal N}=1$ vector multiplet -photino- and three Weyl spinors each one in an ${\mathcal N}=1$ chiral multiplet -"matter fermions"-. The coupling to the 3-form flux corresponds to mass terms for the photino and matter fermions, as well as a photino-matter fermion interaction. The component of the 3-form flux that gives mass to the photino, the $(3,0)$, as well as the component that appears in the photino-matter fermion interaction -nonprimitive $(2,1)$- break supersymmetry in the 10-dimensional IIB background. We saw that in 4-dimensional supersymmetric theories, these interactions come from F and D-term supersymmetry breaking. On the other hand, the mass matrix for the matter fermions, which comes from the $(1,2)$ piece of $G_{(3)}$, can be nonzero in a supersymmetric background corresponding to a D3/D7-brane system plus a 3-form flux perturbation. In the case of a background corresponding only to N D3-branes plus a 3-form flux perturbation, which includes the Polchinski-Strassler solution \cite{PS}, this vev was found in \cite{GP1} to be proportional to the second derivative of the superpotential that breaks ${\mathcal N}=4$ into ${\mathcal N}=1$ in the field theory. These two pieces fit nicely, since in supersymmetric 4-d gauge theories the mass matrix for the fermions in the chiral multiplet is given by the second derivative of the superpotential.

When there is a nonzero gauge field on the D3-brane, we showed that there is a flavor changing dipole moment proportional to the antiholomorphic derivative of the dilaton-axion, as well as a matter fermion-photino dipole moment proportional to its holomorphic derivative. In a IIB supersymmetric background with 4 supercharges, the dilaton is holomorphic or antiholomorphic depending whether we have D3 or anti D3-branes. In the cases considered before, the dilaton is holomorphic, and only the matter fermion-photino dipole moment survives in these supersymmetric backgrounds.

The results here obtained completely agree with the ones obtained in \cite{IIAgaugecompl} for a D0-brane, after performing T-duality in three directions. They provide a very useful result for brane-world models and in general for D-brane phenomenology. The generalization to several D3-branes shouldn't be hard to obtain, and it can be used for the construction of realistic models of nature built from D-branes in background fluxes.     

\section*{Acknowledgements}

I am greatly indebted to Joe Polchinski, for guiding me through this project and reading the manuscript. I would also like to thank Iosif Bena, Andrew Frey, Mark Van Raamsdonk, Pedro Silva, Andrei Starinets, Wati Taylor, Arkady Tseytlin and the CIT-USC Center for Theoretical Physics, where part of this project was completed. This work was supported by National Science Foundation grant PHY97-22022.  

\section*{Appendix}

Throughout the paper, we use two different conventions with respect to indices, according to what's most convenient in each case. In Sections 2,3 and 6, where we deal with ordinary space-time, upper case Latin indices refer to the entire ten-dimensional space-time, and lower case Greek indices refer to 4-dimensional space-time. For the 6-dimensional space, we use m,n,.. when dealing with a real basis, and i,j,... for a complex basis. In Sections 4 and 5, when working in superspace, upper case Latin indices refer to any index, bosonic or fermionic. Lower case Greek indices are fermionic indices and lower case m,n,... refer to bosonic coordinate basis indices; a,b,... refer to bosonic tangent space indices and i,j... are bosonic world-volume indices.
    
We use 32-component Majorana-Weyl spinors $\theta^{\mu}$, where the index $\mu$ should be viewed as a composite index, product of a Majorana-Weyl index and an SO(2) index. The generators of the SO(2) may taken to be $\sigma^1, i\sigma^2, \sigma^3$, where $\sigma$'s are Pauli matrices. A possible correspondence with the complex formalism used in supergravity papers is $\sigma^1 \theta \rightarrow i \theta^*$, $i\sigma^2 \theta \rightarrow -i \theta$, $\sigma^3 \theta \rightarrow \theta^*$. The Dirac gamma-matrices are also extended, and they obey $\{\Gamma^M,\Gamma^N \}=2g^{MN}$ times a delta in the SO(2) index. All fermions and Gamma matrices are thus real.

The $\Gamma$'s may be decomposed
\begin{equation}
\Gamma^{\mu}=\gamma^\mu \otimes 1 \, , \qquad\    \Gamma^{m}= \gamma_{(5)} \otimes \gamma^m \nonumber
\end{equation}
where $\gamma^{\mu}$ and $\gamma^m$ are Dirac gamma matrices corresponding to $SO(3,1)$ and $SO(6)$ respective- ly. The 4-dimensional chirality matrix is
\begin{equation}
\gamma_{(5)}=\frac{i}{4!}\epsilon_{\mu\nu\lambda\rho}\gamma^{\mu\nu\lambda\rho} \nonumber
\end{equation}
where $\epsilon_{0123}=\sqrt{-g_{(4)}}=Z^{-1}$. In the 6-dimensional space, the chirality matrix is
\begin{equation}
\gamma_{(7)}=-\frac{i}{6!}\epsilon_{mnopqr}\gamma^{mnopqr} \nonumber
\end{equation}
and
\begin{equation}
\gamma_{(11)}=\gamma_{(5)}\gamma_{(7)} \nonumber
\end{equation}

The supersymmetry transformations for the fields in type IIB in the string frame are \cite{Schwarz},\cite{HO}
\begin{eqnarray} 
\delta\lambda&=& \frac{1}{2} \Gamma^M \partial_M \phi \varepsilon  -\frac{1}{24}\Gamma^{MNP} H_{MNP}\, \sigma^3\, \varepsilon - \frac{1}{2}e^{\phi}\,  
\Gamma^{M}F_{M}\, (i\sigma^2)\, \varepsilon - \frac{1}{24}e^{\phi}  
\Gamma^{MNP}F'_{MNP}\, \sigma^1\, \varepsilon \nonumber\\
\delta\psi_{M}&=&  D_{M} \varepsilon +\frac{1}{8} e^{\phi}\,\Gamma^N \Gamma_M\, F_N \,(i\sigma^2) \varepsilon - \frac{1}{8} 
\Gamma^{PQ} H_{MPQ}\, \sigma^3 \varepsilon + \frac{1}{48} e^{\phi}\, 
\Gamma^{PQR} \Gamma_M \, F'_{PQR}\, \sigma^1 \varepsilon \nonumber \\
& & + \frac{1}{16 \cdot \,5!} e^{\phi}\, 
\Gamma^{PQRST} F'_{PQRST}\, \Gamma_M\, (i\sigma^2) \varepsilon 
\nonumber\\
\delta e_m^a&=& \overline{\varepsilon}\Gamma^a \psi_m \nonumber\\
\delta \phi&=& \overline{\varepsilon}\lambda \nonumber\\
\delta B_{mn}&=& 2 \overline{\varepsilon} \sigma^3 \Gamma_{[m} \psi_{n]} \nonumber\\
\delta C_{mn}&=& - 2 e^{-\phi}\, \overline{\varepsilon}\, \sigma^1 \Gamma_{[m} \left(\psi_{n]}-\frac{1}{2} \Gamma_{n]} \lambda\right)+ C \delta B_{mn} \nonumber\\
\delta C_{mnop}&=& -4 e^{-\phi}\,\overline{\varepsilon}(i\sigma^2)\Gamma_{[mno}\left(\psi_{p]}-\frac{1}{4} \Gamma_{p]} \lambda\right) + 12\, C_{[mn} \delta B_{op]} \nonumber
\end{eqnarray}
where $D_M$ is the covariant derivative with respect to the metric, and the modified field strengths $F'$ are
\begin{equation}
F'_{(3)}= dC_{(2)} - C\, H_{(3)} \, , \qquad F'_{(5)}= dC_{(4)}-H_{(3)} \wedge C_{(2)}
\end{equation}

The commutator of two local supersymmetries on a gauge field gives a combination of a coordinate and a gauge transformation. For any field to which the commutator is applied, the resulting coordinate transformation is
\begin{equation}
\zeta^m= \overline{\varepsilon}_2\Gamma^m\varepsilon_1
\end{equation}
and the gauge transformations are
\begin{equation}
\Lambda^B_m= \zeta^n B_{mn} - \overline{\varepsilon}_2\, \sigma^3 \Gamma_m \varepsilon_1
\end{equation}
for the NS-NS 2-form field $B_{(2)}$, and
\begin{eqnarray}
\Lambda^C_m&=& \zeta^n C+ e^{-\phi} \overline{\varepsilon}_2 \,\sigma^1\, \Gamma_m \varepsilon_1 - C \overline{\varepsilon}_2\,\sigma^3 \Gamma_m \varepsilon_1\nonumber\\
\Lambda^C_{mnp}&=& \zeta^q C_{qmnp}+ \frac{1}{2}e^{-\phi}\, \overline{\varepsilon}_2\, (i\sigma^2) \Gamma_{mnp} \varepsilon_1 - 3\, C_{[mn} \overline{\varepsilon}_2\,\sigma^3 \Gamma_{p]} \varepsilon_1
\end{eqnarray}
for the R-R 2- and 4-form fields $C_{(2)}$ and $C_{(4)}$ respectively.

\end{document}